# SISTEMA DE MONITOREO DE VARIABLES AMBIENTALES CON TECNOLOGÍA IoT

Harold Pinilla [1]
*hpinilla@academia.usbbog.edu.co*,

José Macias [2]
*jmacias@academia.usbbog.edu.co*,

Emmanuel Lescano [3]
*electronica.practica@usbbog.edu.co*,

José David Alvarado [4]
*jalvarado@usbbog.edu.co*,

Wilder Castellanos [5]
*wcastellanos@usbbog.edu.co*,

[1] Estudiante Ingeniería Electrónica, Universidad de San Buenaventura, Colombia

[2] Estudiante Ingeniería Electrónica, Universidad de San Buenaventura, Colombia

[3] Estudiante de Ingeniería Electrónica, Universidad Tecnológica Nacional, Argentina

[4] Magister en Ingeniería, Universidad de Ibagué, Colombia. Profesor Asociado: Universidad de San Buenaventura, Colombia

[5] Doctor en Telecomunicaciones, Universidad Politécnica de Valencia, España. Profesor Titular: Universidad de San Buenaventura, Colombia

***Resumen* — Este artículo describe el diseño de una plataforma de monitoreo de variables ambientales aplicado a la agricultura, flexible y de bajo costo. Para la construcción de esta plataforma se utilizaron tecnologías basadas en protocolo de comunicación, Wifi, Bluetooth, Zigbee, utilizando sistema embebido Raspberry pi 3 b+ y sensores para cuantificar diferentes variables ambientales, mediante diferentes herramientas hardware y software de código abierto. La red está compuesta por un nodo central (gateway), implementado sobre la tarjeta Artik 1020 de Samsung, y dos nodos donde se encuentran conectados los sensores para las lecturas de las variables ambientales. Finalmente, los datos son recolectados por el gateway, el cual se encargará del procesamiento y almacenamiento en una base de datos para que el usuario en un futuro pueda acceder a la información en tiempo real desde cualquier lugar.**

# INTRODUCCIÓN

El crecimiento económico en la actualidad afecta directamente a el cambio climático, por este fenómeno ha influido en aspectos como; la temperatura atmosférica, la variabilidad en el régimen de lluvias, la duración de las sequias modificaciones en los patrones climáticos, entre otros [1] . Los efectos negativos ocasionados afectan a la población a nivel mundial, por ejemplo, este fenómeno se puede evidenciar en la agricultura en donde los cambios climáticos ocasionan diminución o perdida de las cosechas, lo cual afecta negativamente la economía de los países no desarrollados debido a que es una de las actividades económicas principales y además en necesaria para satisfacer la demanda de alimento de la población a nivel mundial [2].

Para contrarrestar los efectos del cambio climático en la agricultura, podemos encontrar diferentes alternativas como los sistemas para monitorear las diferentes variables necesarias para establecer el comportamiento del clima y a partir de esta información recolectada construir modelos de los posibles patrones climáticos y de esta manera poder realizar las acciones preventivas para que no se pierdan las cosechas [3], o diferentes sistemas de monitoreo de variables [4], [5], o soluciones aplicando internet de la cosas [6], [7]. Para lograr construir estos modelos es necesario contar con diferentes estaciones de recolección de la información dependiendo de la extensión del cultivo o el sector en donde se desea implementar este tipo de sistema, para lo que resulta conveniente implementar un sistema basado en el modelo del internet de la cosas (IoT) [8], [9], en donde la información recopilada por cada estación debe estar almacenada en una base de datos alojada en la nube, lo que permite que la información se pueda acceder de diferentes herramientas y además es un modelo en el cual es posible ajustar el número de estaciones dependiendo de las características del entorno.

En términos de hardware los dispositivos para implementar un sistema de monitoreo de variables deben ser sistemas embebidos que incorporen los componentes necesarios para una aplicación de IoT [10], además resulta adecuado la posibilidad de poder utilizar diferentes protocolos de comunicación inalámbrica para poder seleccionar el que se adecue mejor a la características del entorno, la aplicación y los requerimientos del sistema, por último es necesario los sensores para realizar la medición de las variables a monitorear.

Es así que este proyecto se enfocara en el diseño de una red de sensores inalámbrica, compuesta por un Gateway y nodos inalámbricos utilizando protocolos de comunicación wifi, bluetooth y zigbee, aplicando el modelo del internet de la cosas, en donde se realizara la medición de variables como; temperatura, humedad, velocidad y dirección del viento, radiación solar y precipitación, para que esta información esté disponible para en un futuro los agricultores puedan analizar los datos y tomar las medidas requeridas para mantener los cultivos adecuadamente; estos datos serán obtenidos por medio de un microcontrolador Raspberri PI 3 en donde estarán conectados módulos de sensores y de comunicación inalámbrica, y por último los datos serán enviados a una tarjeta ARTIK 1020 que incorporara las funciones de un Gateway para IoT, la que estará encargada del procesamiento, control y almacenamiento en una base de datos.

# DESARROLLO

## *A.* ***Diseño de la plataforma meteorológica***

En sistema de monitoreo de variables los nodos están desarrollados en una raspberry pi 3 b+, en donde cada uno tiene asociado los sensores para realizar la medición de seis variables meteorológicas distintas: humedad relativa; temperatura ambiente; radiación solar; nivel de precipitaciones; dirección y velocidad del viento. Para realizar la trasmisión de comunicación cada nodo incorporar la posibilidad de relazar la conectividad inalámbrica utilizando los protocolos; wifi, bluetooth y Zigbee, pero solo es posible utilizar un solo protocolo en cada estación.

Por el lado del Gateway IoT, está diseñado en el sistema embebido Samsung Artik 1020, en donde se recopila la información de los diferentes nodos utilizando el protocolo MQTT, además incorporar los mismos protocolos de conectividad inalámbrica de los nodos, pero en este caso el Gateway está configurado para realizar conexión bidireccional con cada nodo utilizando todos los protocolos de manera simultánea, además realizar la suscripción de la información un servidor público utilizando internet, se definió una topología de red tipo estrella.

En la **FIG. 1** se muestra el diagrama general de los componentes del sistema de monitoreo de variables.

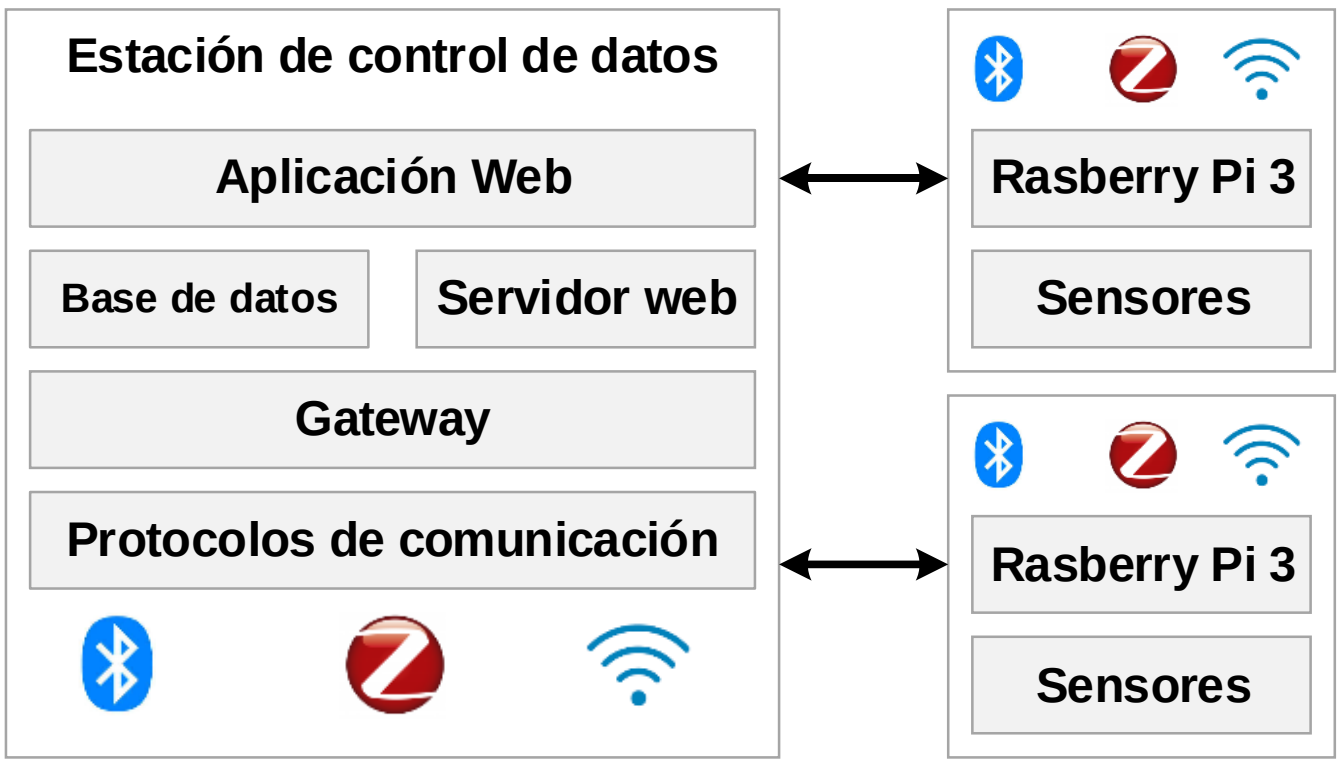

*FIG. 1* - Estación de control de datos.

## B. *Gateway IoT*

Como Gateway IoT se utiliza el kit de desarrollo Samsung Artik 1020, es una tarjeta de desarrollo de alto rendimiento, multiprotocolo ya que cuenta con la posibilidad de comunicarse inalámbricamente mediante Bluetooth, ZigBee y Wi-Fi [11], además de contar con múltiples puertos de I/O capaz comunicarse mediante los módulos I2C, SPI, UART, entre otros.

En la **FIG. 2** se ilustran las características de esta placa de desarrollo, así como también sus protocolos de comunicación y características de hardware.

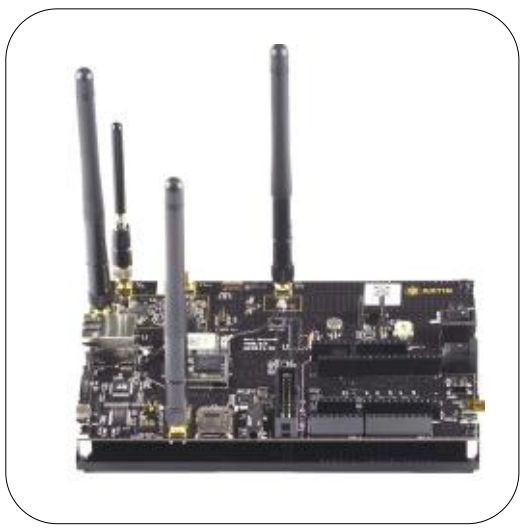

| Media |
|---|
| Camera I / F: 1x 2-lane MIPI CSI up to 3 M P at 30 fps / 1x 4 - lane MIPI CSI up to 16 MP at 30 fps |
| Display: 4 - lane MIPI DSI up to FHD / 1920 x 1200 at 24 bpp simultaneous HDMI / 1920 x 1080 at 60 fps |
| Audio: 1x channel PCM, 2-channel I²S |

| Conectivity |
|---|
| Wifi IEEE 802.11a/b/g/n/ ac |
| LPWA |
| Bluetooth 4.1 + smart |
| Zigbee |
| Z-wave |

| Processor |
|---|
| CPU: quad core Cortex® - A15 at 1.5 GHz, and quad core Cortex® - A7 at 1.3 GHz |
| GPU: Mali™ - T628 MP6 |

| Interfaces | | |
|---|---|---|
| ADC | UART | GPIO |
| JTAG | $I^2C$ | SDIO |
| USB | SPI | $I^2S$ |

| Memory |
|---|
| DRAM: 2 GB LPDDR3 |
| Flash: 16 GB eMMC |

*FIG. 2* - *Características* de la tarjeta de desarrollo Samsung Artik 1020

## C. *Nodo IoT*

Cada nodo IoT inalámbrico dispone de los sensores de temperatura ambiente y humedad relativa, am 2315; radiación solar, Davis 6450; nivel de precipitaciones, dirección del viento y velocidad del viento, SEN-08942; luego de un acondicionamiento de la señal proveniente de cada uno de estos sensores, se conectan a un sistema embebido Raspberry pi 3 b+ el cual es el encargado de procesar los datos, cuantificarlos y enviarlos al Gateway central. El paquete de datos se envía mediante los protocolos Bluetooth, Zigbee o Wifi, según se requiera. En la *FIG. 3* se presenta la arquitectura utilizada en cada nodo IoT, con sus respectivas características de comunicación, librerías utilizadas, y hardware.

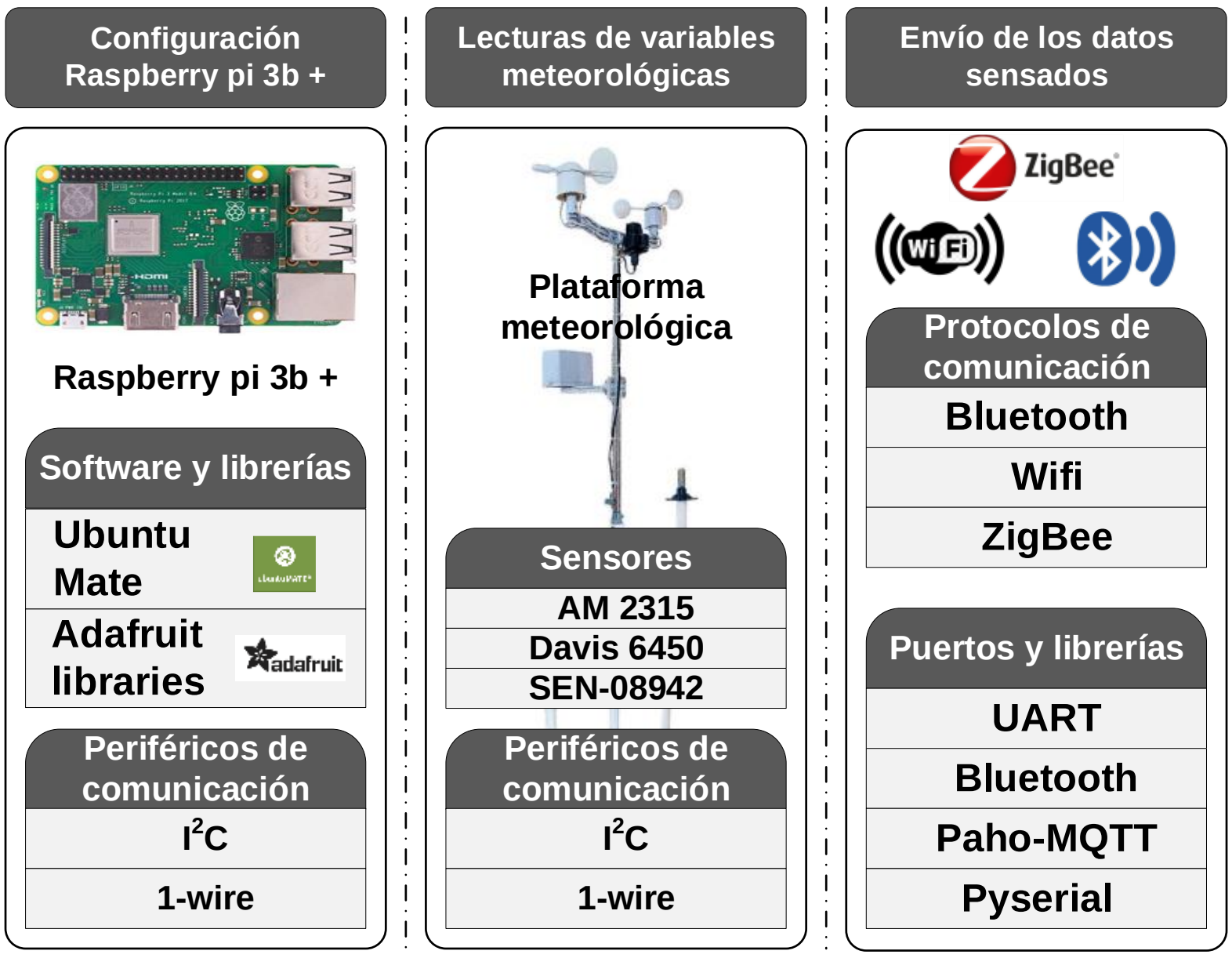

*FIG. 3 - Descripción Nodo IoT*

Como unidad central de procesos de cada nodo IoT, se utiliza la ya mencionada tarjeta embebida Raspberry pi 3 b+, esta tarjeta es básicamente un minicomputador que funciona con software de código abierto [12], la cual tiene conexiones inalámbricas como Wi-Fi y Bluetooth [13], además cuenta con la posibilidad de utilizar puertos de I/O, y la aplicación de múltiples protocolos de comunicación como son UART, I2C, SPI, etc.

En la *Fig. 4* se ilustran las características de esta placa de desarrollo, así como también sus protocolos de comunicación y características de hardware.

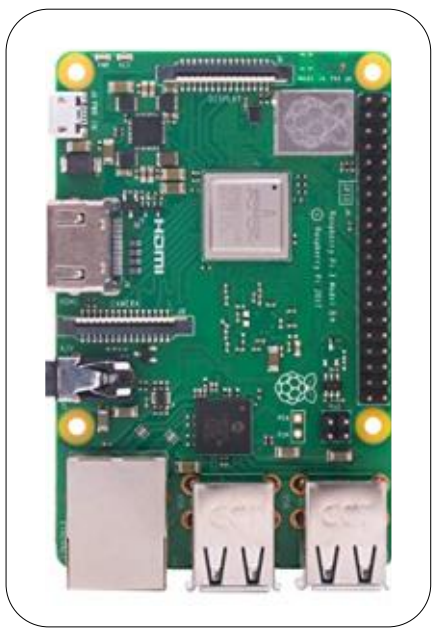

| Media |
|---|
| 1 × full size HDMI |
| MIPI DSI display port |
| MIPI CSI camera port |
| 4 pole stereo output |
| Composite video port |
| H.264, MPEG-4; OpenGL ES 1.1. |

| Puertos y librerías |
|---|
| GPIO |
| UART |
| $I^2C$ |
| SPI |
| USB 2.0 |
| SDIO |

| Conectivity |
|---|
| 802.11.b/g/n/ac , BT 4.2, BLE |
| Gigabit Ethernet over USB 2.0 |
| 4 × USB 2.0 ports |

| Processor |
|---|
| BCM2837B0, Cortex-A53 64-bit @ 1.4GHz |

| Memory |
|---|
| 1GB LPDDR2 SDRAM |

*Fig. 4 - Características tarjeta Raspberry pi 3 b+.*

### D. *Sensores utilizados*

El sensor utilizado para la medición de la temperatura ambiente y humedad relativa es el am2315. Este cuenta con una precisión en la medición de humedad de ±2% y en la medición de temperatura de ±0.1°C. La comunicación con el sistema embebido Raspberry pi es a través del protocolo estándar $I^2C$ [14]. En la **Fig. 5** se muestra el sensor mencionado.

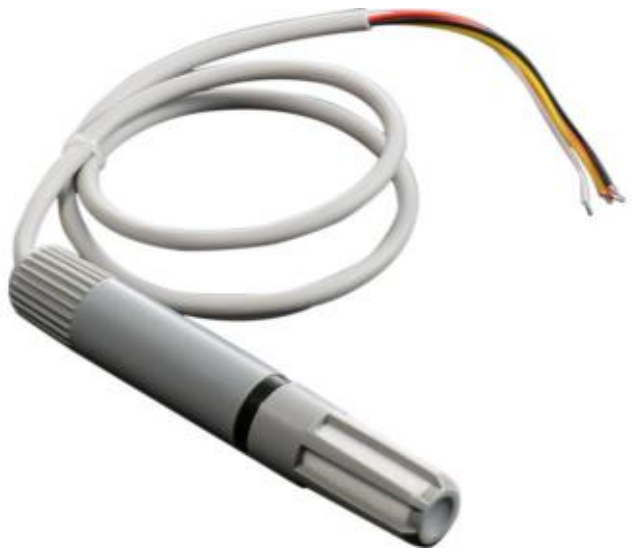

*Fig. 5* - Sensor de medición de temperatura ambiente y humedad relativa AM2315.

Por el lado de la radiación solar, el sensor Davis 6450 es el encargado de realizar la medición, este transductor ofrece una salida analógica de 0 a 3Vdc, con una resolución de 1,67mV por W/m2 [15]. En la *Fig. 6* se expone el sensor Davis 6450.

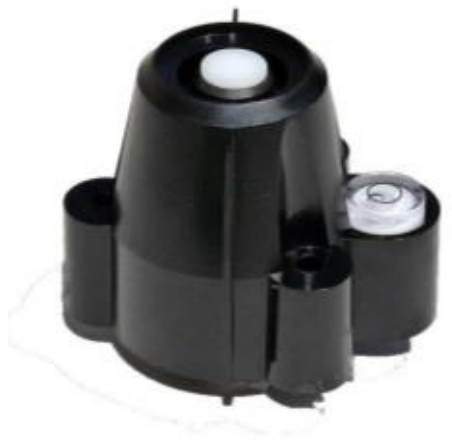

*Fig. 6 - Sensor de radiación solar* ***Davis 6450***

Por último, el kit de medición SEN-08942, el cual cuenta con tres tipos diferentes de sensores para las mediciones de nivel de precipitaciones, dirección y velocidad del viento. El sensor para la medición del nivel de precipitaciones funciona accionando un switch cada 0.011'' de agua de lluvia recolectados en una cubeta. Para la medición de la velocidad del viento se emplea un anemómetro, un switch equipado con un encoder se activa por cada rotación de las aspas, finalizando se dispone para la medición de la dirección de viento una combinación de resistencias, para las cuales midiendo el voltaje de salida se determina tal magnitud meteorológica, cumpliendo la función de una veleta [16]. El kit de medición SEN-08942 se expone a continuación en la *Fig. 7*.

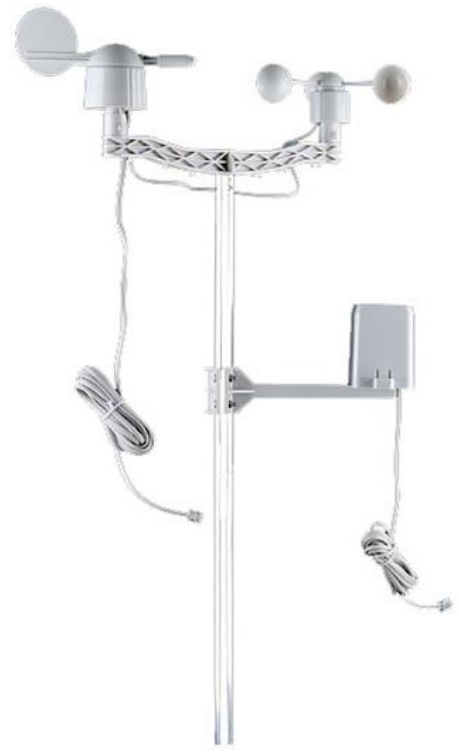

*Fig. 7 – Kit de medición* ***SEN-08942***

La tarjeta Raspberry Pi no incluye un módulo ADC, se anexa el CI mcp3008, como complemento para los sensores en los que su respuesta de salida es analógica.

La medición de los diferentes sensores se gestiona bajo la tarjeta embebida Raspberry pi 3b +, en esta se programan las diferentes rutinas de inicialización, medición y cuantificación de los datos. En la *FIG. 8* se muestra un breve resumen del procedimiento seguido para la operación de los sensores.

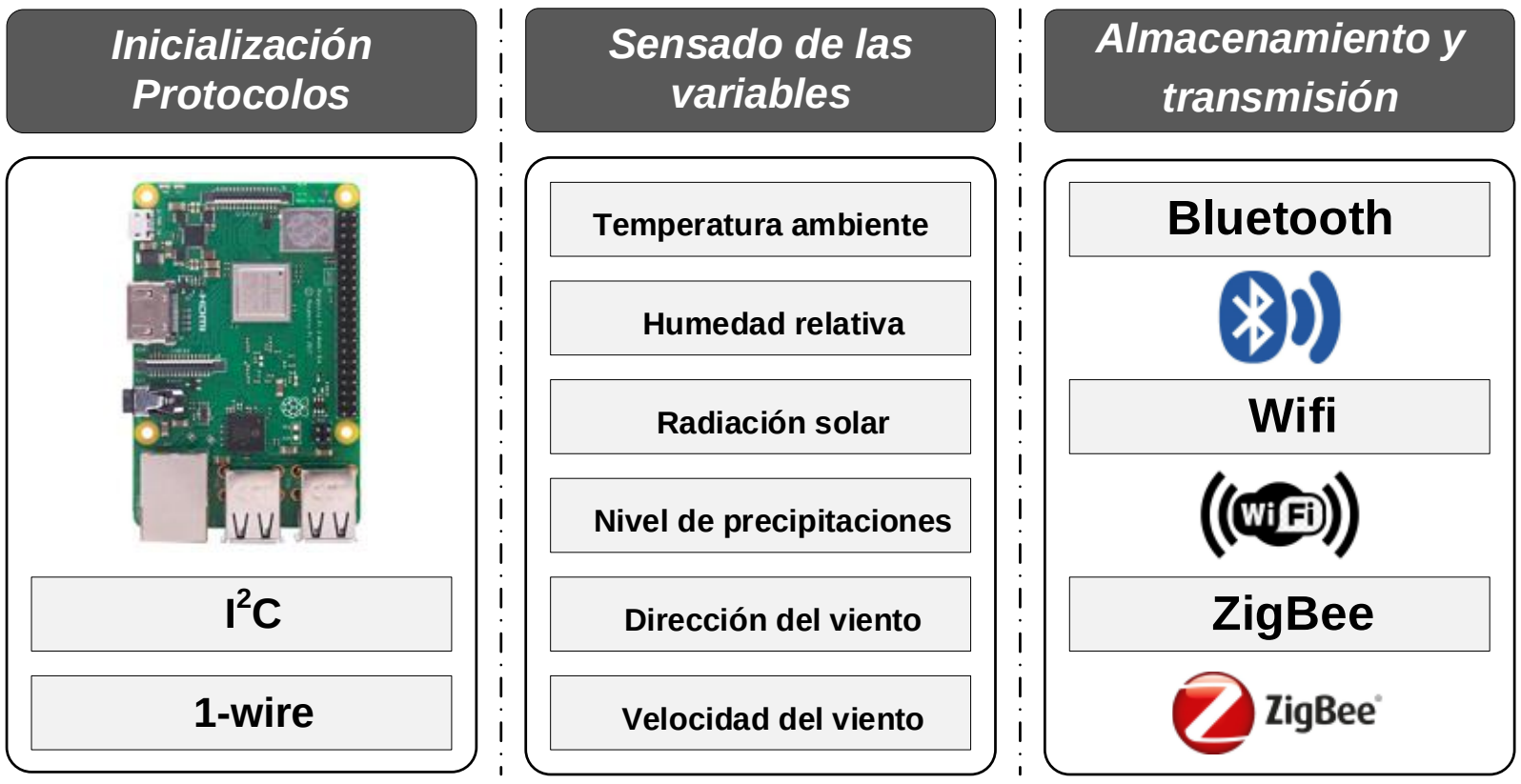

*FIG. 8 - Diagrama de monitoreo de las variables meteorológicas*

## ANÁLISIS DE RESULTADOS

En las siguientes figuras se ilustran las gráficas de temperatura, humedad, velocidad del viento y radiación solar respectivamente, medidas por ambos nodos durante aproximadamente 7 minutos, en donde se realizó la captura de 36 muestras.

En la *FIG. 9* se gráfica la temperatura ambiente medida por los sensores AM 2315 de ambos nodos, ubicados en dos puntos cercanos, se evidencia un cambio proporcional en los valores monitoreados y teniendo una diferencia máxima de 0.3 C°.

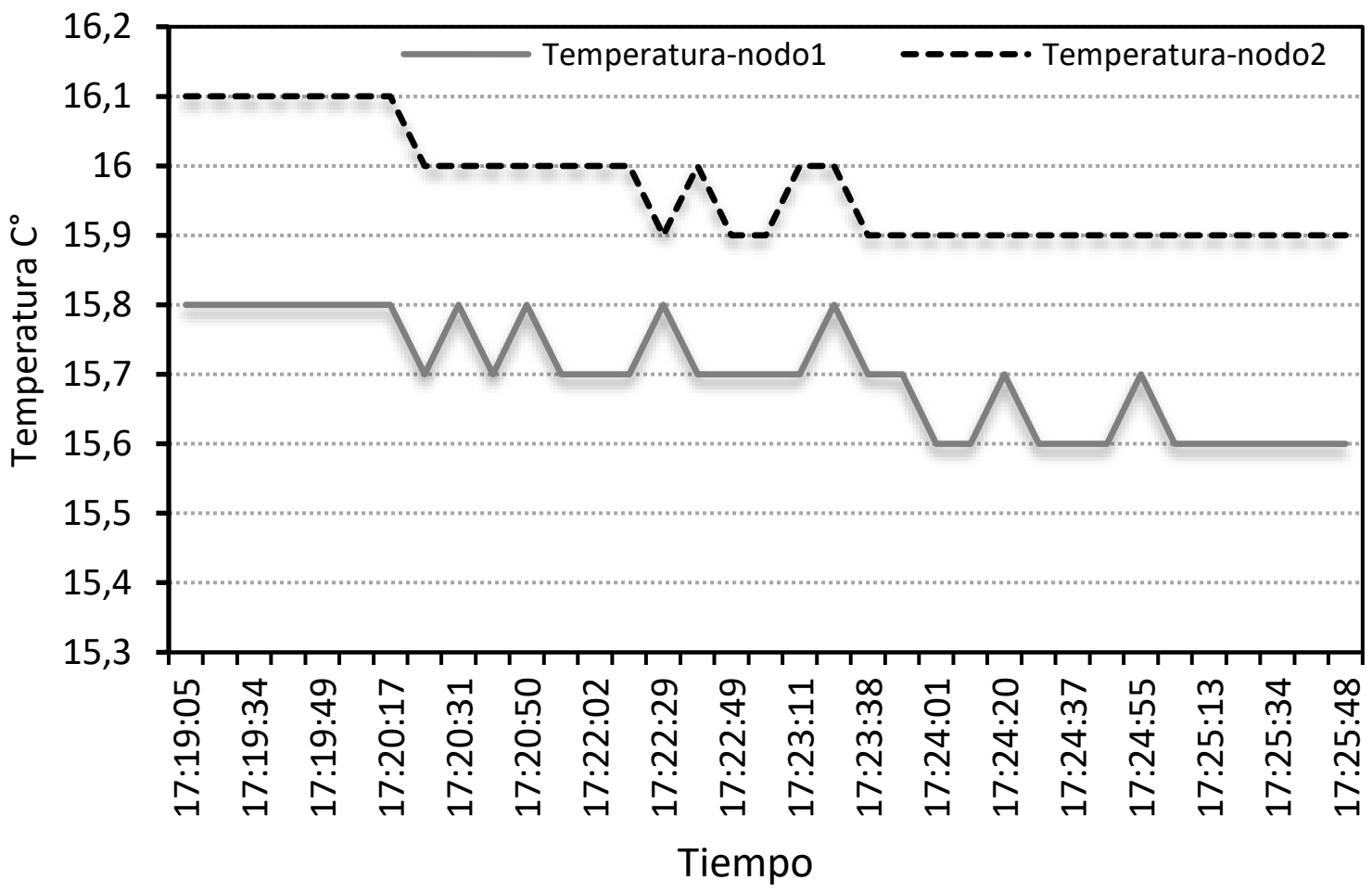

*FIG. 9 - Gráfica de temperatura de los nodos*

En la *FIG. 10* se ilustra la humedad relativa medida por los sensores AM 2315 de ambos nodos IoT, se muestra el cambio en porcentaje de humedad, se observa que la variación de los datos es similar en los dos casos, obteniendo la mayor diferencia en porcentaje de humedad de 1%.

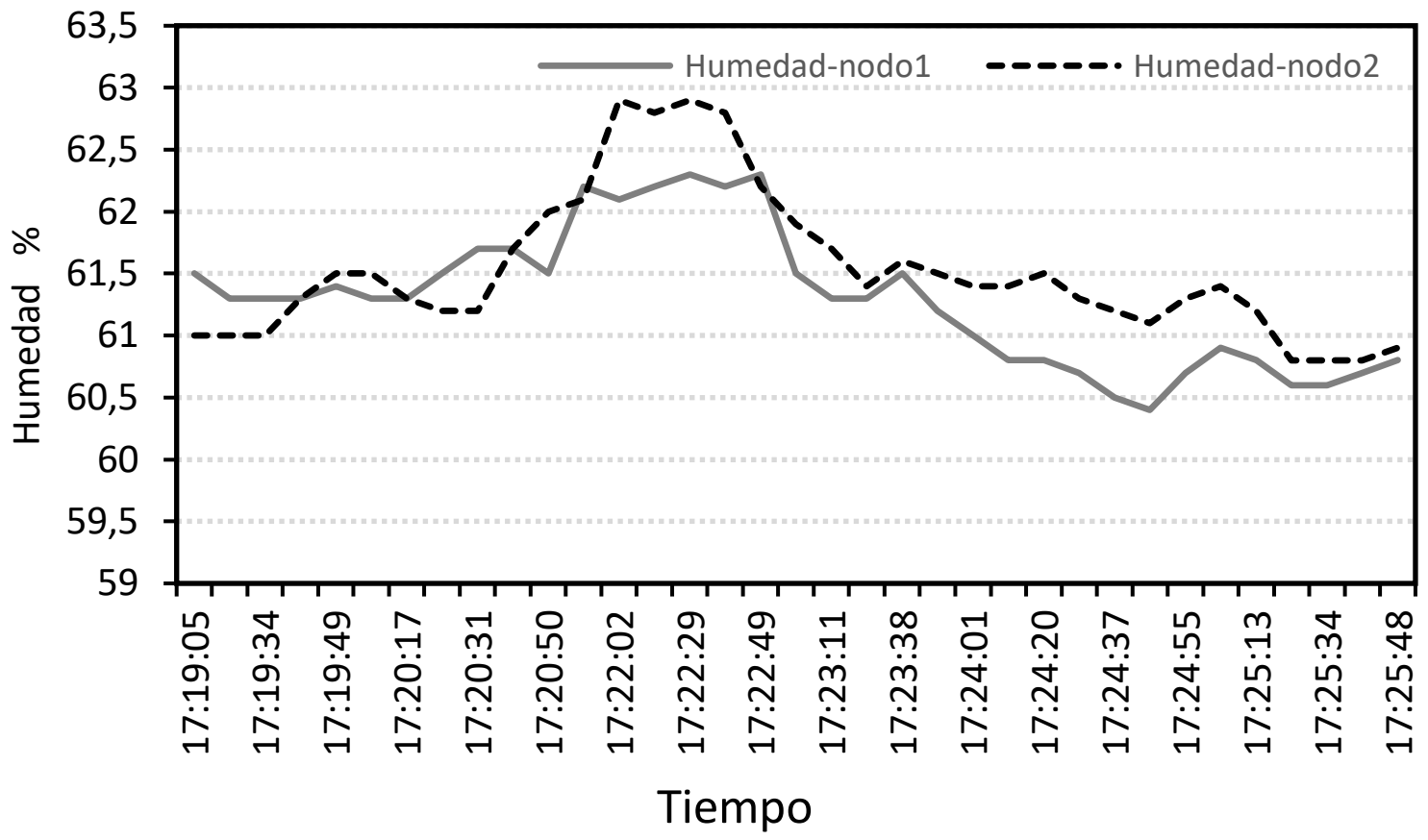

*FIG. 10 - Gráfica de humedad de los sensores*

La *FIG. 11* muestra la velocidad del viento tomada por el kit de medición SEN-08942 en ambos nodos, ubicados en distintos lugares obteniendo el valor de la velocidad de viento en Km/h, dando como resultado una variación similar en los dos nodos.

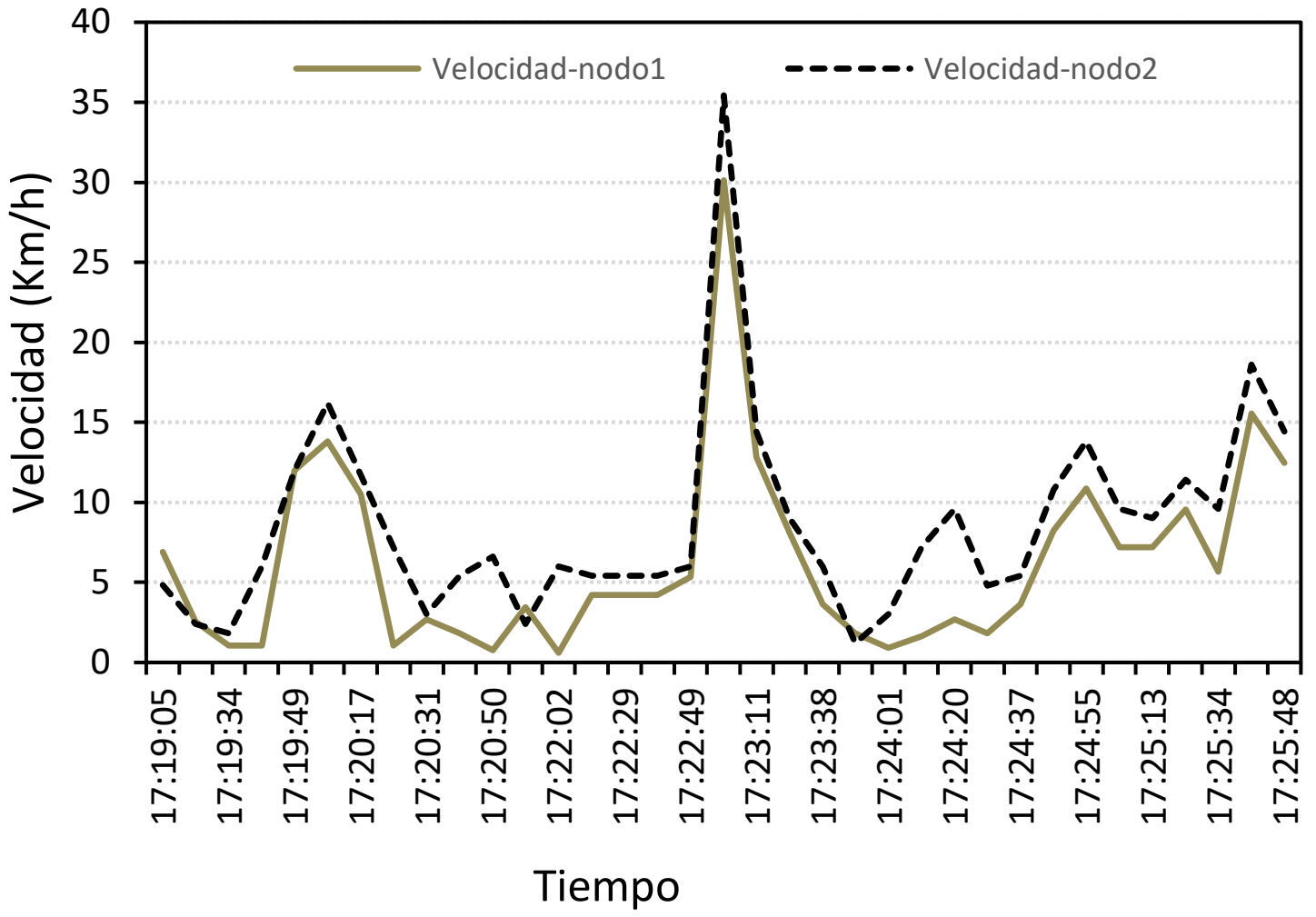

*FIG. 11 - Gráfica de velocidad del viento de los nodos*

La *FIG. 12* expone la radiación solar capturada por el sensor *Davis 6450* por ambos nodos, los datos se generan en unidades W/m$^2$, se evidencia un cambio relacionado de los datos de radiación durante el tiempo de observación.

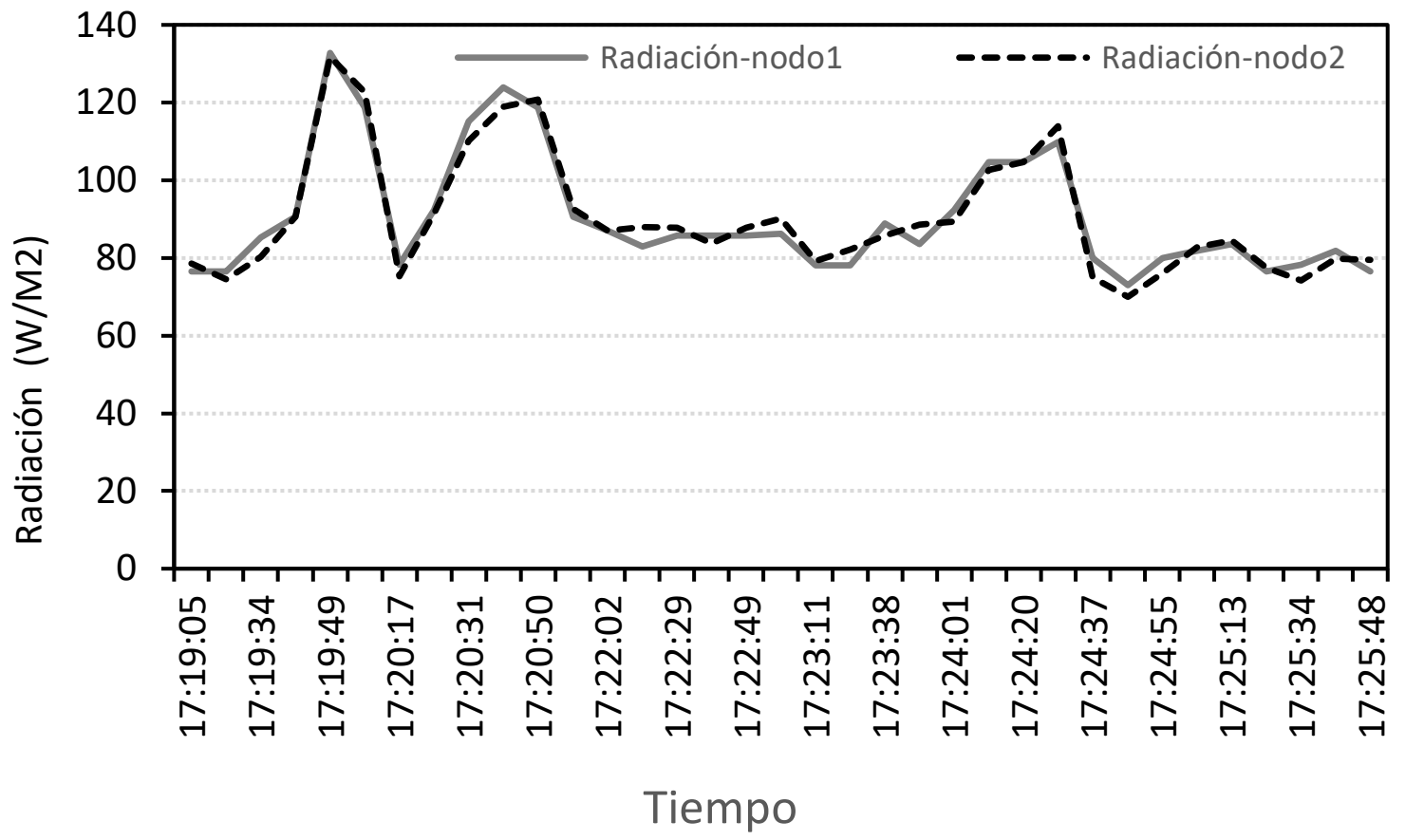

*FIG. 12 - Gráfica de radiación de los sensores*

# CONCLUSIONES

La implementación de los diferentes protocolos de comunicación inalámbrica para los nodos (Rasberry PI) y el Gateway (ARTIK), resultaron adecuados debido a que para Wifi y Bluetooth las tarjetas incorporan de fabrica estos interfaces por lo dispones de diferentes herramientas para realizar las diferentes configuraciones necesarias para la implementación de la red, para el caso de Zigbee fue necesario adicionar un módulo USB con este protocolo y por medio de la interfaz de comunicación serial fue posible incluirlo en los nodos y el Gateway.

Los sistemas operativos de las diferentes tarjetas utilizadas son basados en diferentes distribuciones de Linux, y para cuales fue posible realizar la instalación de Mosquitto-MQTT con el que cual se presentó una completa compatibilidad, a partir de esta herramienta de mensajería para el internet de las cosas fue posible la implementación del sistema de intercambio de información basados en el modelo de publicador/suscritor para la red de sensores implementada.

Los nodos recopilaron adecuadamente la información de los diferentes sensores utilizados a una frecuencia establecida por el usuario, para lo que fue necesario el diseñó de un paquete de datos con el nombre, unidades, valor, hora de captura y protocolo de envío utilizado para cada sensor; con lo que se obtuvo una simplificación en la clasificación, manejo y almacenamiento en el proceso de envió al Gateway.

Se tiene previsto como trabajo futuro, el aumento de los nodos Raspberry pi 3 b + y la configuración de los mismo de manera remota. Así como la implementación de un actuador e incluir un sistema de autonomía energética, utilizando energía fotovoltaica.